\documentclass[12pt, letterpaper] {article}      
\usepackage{osajnl2}
\usepackage{amsmath}		
\begin{document}

\title{Calibration of force actuators on an adaptive secondary prototype}

\author{Davide Ricci$^{1,2,*}$, Armando Riccardi$^{2}$, Daniela Zanotti$^{2}$}

\address{$^{1}$Institut d'Astrophysique et de G\'eophysique, Universit\'e de Li\`ege,
\\ All\'ee du 6 Ao\^ut, B-4000 Li\`ege, Belgium.
\\ $^{2}$INAF - Osservatorio Astrofisico di Arcetri, 
\\ L.go E. Fermi 5, 50125 Firenze, Italy.}

\address{$^*$Corresponding author: ricci@astro.ulg.ac.be}
\begin{abstract}
In the context of the LBT project, we present the results of force actuator calibrations performed on an adaptive secondary prototype called P45: a thin deformable glass, supplied with magnets glued on its back. 
Electromagnetic actuators, controlled in a closed loop with a system of internal metrology based on capacitive sensors, continuously deform its shape to correct the distortions of wavefront.
Calibrations of the force actuators are needed because of the differences between driven forces and measured forces. This paper describes the calibration procedures and the results, obtained with  errors presents less than 1.5\%. 
\end{abstract}
%
\maketitle 
\section{Introduction}
Adaptive optics has been developed as a technique to correct, in ground-based telescopes, the distortions of wavefront due to the atmospheric turbulence\cite{roddier}.
During the past two decades, several solutions were proposed as wavefront correctors, for example post-focal deformable mirrors using piezoelectric actuators and bimorph piezoelectric mirrors. These kinds of solutions are limited to a maximum stroke of a few microns $\approx 5\rm\mu m)$, that implies the necessity to correct the tip-tilt error with a separate mirror.
Techniques based on the principles of electromagnetism exist in the field of astronomy or optometry for human eye applications. These small correctors use a thin membrane coated with a magnetic layer\cite{1998ITM343564D}, a special kind of magnetic liquid\cite{2006OExpr1411486B}, or a coated membrane provided with magnets\cite{Fernandez:06}.
This last solution follows in time the innovative technique based on adaptive secondary mirrors as wavefront correctors, previously proposed\cite{1993_salinari_pisc_48_247} to exceed the limits of the piezoelectric systems. The advantages of the novel technique, with respect to the other systems cited before, are several. A first peculiarity is to be a pre-focal corrector (the telescope's traditional secondary mirror is removed and substituted with the adaptive secondary). Electromagnetic actuators, after reciving a certain current, create a magnetic field  that acts on an equal number of magnets glued on the back of a Zerodur glass mirror, deforming its shape. As there is no contact between the two components, this wavefront corrector has less limits in stroke with respect to the previous techniques, and allows to include in a unique entity both the tip-tilt and high oders correctors. The instrument is supplied with an internal metrology system based on capacitive sensors. Moreover, with respect to a traditional telescope design with an adaptive optics system, using the secondary adaptive prototype means to avoid reflections and transmissions, with an increase of efficiency. Finally, the choice of Zerodur glass avoids any hysteresis effect\cite{2002bcaoconf55R}.
This technique was successfully applied\cite{2003_brusa_spie_5169_26}: after the construction of two smaller prototypes, a 336 actuators adaptive secondary mirror called MMT336 was realized. At present, the instrument is fully working on the MMT telescope, and it is the first and only adaptive secondary mirror operating on a telescope.
Furthermore, following the positive experience of the MMT336 and its prototypes, two 672 actuators adaptive secondary mirrors for the LBT telescope are being developed\cite{2003_riccardi_spie_4839_721}. A first  prototype called P45, that includes new solutions for the mirror shape, the control electronics and force actuators, has recently ended its test study phase\cite{2004_riccardi_spie_5490_1564}.

This paper describes the procedure of calibration of the force actuators of P45. These calibrations, not performed before on the existing adaptive secondary mirrors, allow the increase of the determination of the stiffness modes of the mirror, in order to obtain a base that can be used in the modal optical loop.
A brief description of the P45 prototype is provided in Section \ref{julia}. Then, the linear systems used to find the calibration are treated  in Section \ref{sis}, and Section \ref{proof}  explains the proof code. Finally, laboratory results and conclusions  are presented in Sections \ref{lab} and \ref{fine}.
\section{The P45 prototype}
\label{julia}
The P45 prototype of the LBT adaptive secondary mirror consists of a three layer structure (see fig.~\ref{p45}): a thin deformable shell; a thick reference plate; and a third plate that acts as an actuator support and heat sink.
The thin ($1.6\rm mm$) \emph{deformable shell}, realized in Zerodur Glass, is supplied with 45 magnets glued on its back with a 120 degrees symmetry, and arranged in 3 concentric rings (see fig.~\ref{Geometria}). On the surfaces of the shell, two thin aluminium films are applied, one on each side.
A thick Zerodur glass plate, named the \emph{reference plate}, provides a reference surface for the back side of the thin shell. This plate is supplied with holes in which are inserted, from the bottom, the electromagnetic actuators.
Each magnet of the thin shell faces the coil of the respective actuator, provided with a capacitive sensor capable to act as an internal metrology system. Currents driven on the coils of the actuators generate an electromagnetic force pattern that deforms the mirror acting on the magnets. Inside each actuator is placed a security magnet, called \emph{bias magnet}, which is used to sustain the mirror in case of a current blackout\cite{2004_riccardi_spie_5490_1564}. The bias magnets are a new feature of the P45 with respect to the previous adaptive secondary mirrors.
Finally, the heat sink of the actuators is assured by a thick \emph{cold plate}, to which the actuators are fixed.

This system requires calibration because of the differences between the driven force on the
actuators and the measured force by the control electronics.
Force calibrations allow to increase the precision for the identification of the natural modes (stiffness modes) of the mirror, in order to obtain a  base that can be used in the modal optical loop to build the force patterns to compensate for the wavefront error. In particular, they allow to determine the maximum number of correcting modes for a given dissipation threshold of the actuators, with respect to other modal bases.\cite{2006_riccardi_spie_6272_4}
\section{The linear systems}
\label{sis}
The actuators of the P45 deformable glass provide the needed currents to move the mirror acting on the magnets, by introducing forces orthogonal to its surface. 
In order to obtain the equilibrium, at a given distance from the reference plate (typically $68\rm\mu m$), the cardinal equations of statics must be satisfied: the sum of forces and torques driven to each actuator must be equal and opposite to the sum of the other forces and torques acting on the mirror.
These last ones are due to the weight force $P=(-2.920\pm 5\times 10^{-3}) \rm N$ and the total force of bias magnets $B$, unknown. The intersection between the shell plane and the optical axis $z$ of the thin shell is chosen as the centre of resolution for torques, while the $x$ and $y$ axes are chosen as in fig. \ref{Geometria}.
In this configuration, and because forces are introduced orthogonally to the mirror surface, the torque on the $z$ axis can be disregarded.
So the considered torques are those through the $x$ and $y$ axes, named respectively $M_y$ and $M_x$ (including the torque of the bias magnets), and the sign of $M_y$ will be negative for the right-hand rule.
The balance of the force and torque must be satisfied in each position and for each shape that the mirror will take. If $N$ corresponds to the number of these shapes ($N\ge 45$), $i=1\dots 45$ represents the actuator index and $f$ are the forces:
\begin{eqnarray}
\label{equ}
\label{prima}
				\ -\sum_{i}\  f^{i}_{1} =\ P + B 				\hskip 1 truecm
				&\cdots& 											\hskip 1 truecm
				-\sum_{i}\ f^{i}_{N} = P + B  				\nonumber \\
				-\sum_{i} x^{i}f^{i}_{1} =\ M_y \			\hskip 1 truecm
				&\cdots& 											\hskip 1 truecm
				-\sum_{i} x^{i}f^{i}_{N} = M_y \ 			\\
				-\sum_{i} y^{i}f^{i}_{1} = -M_x 				\hskip 1 truecm
				&\cdots& 											\hskip 1 truecm
				-\sum_{i} y^{i}f^{i}_{N} = -M_x \ 			\nonumber 
\end{eqnarray}
The bias magnet force $B$ will be positive because in the opposite direction with respect to the weight force of the mirror.
To set the forces that are solutions of the previous equations we need to supply to the actuator a certain current, multiplied by an adequate calibration constant $\alpha$, most likely different for each actuator:
\begin{equation}
f^i_{N}=\alpha^i c^i_{N}\ \ .	
\label{seconda}
\end{equation}
Being in a closed loop system, the control electronics measures differences of force that fluctuate around a zero value. For this matter we can treat this process in the linear regime to write the relation that impose the stability.
The $c^i_N$ are the forces driven to the actuators by the control electronics, that we call \emph{commanded forces}; and the $f^i_N$ are the forces really measured by the capacitive sensors, that we will call emph{true forces}. The calibration constants $\alpha^i$ are dimensionless: the ``true~Newton~/~commanded~Newton'' unit of measurement is used.  The $\alpha^i$ are unknowns.
The following section will illustrate the method used to set the calibrations.
The requested driven force sets are obtained as follows: the mirror is deformed driving each actuator in a determined position; the shape is applied from the closed loop system,  that automatically controls the forces to keep the mirror stable. Such forces are recorded by the control electronics and used to build the $c^i$ vectors.
Combining eqs. \ref{prima} and \ref{seconda}, to impose the stability means to find the $\alpha^i$ that are solutions of the following linear system:
\begin{equation}
-
\begin{pmatrix}
	c_1^1 		 &		c_1^2 			& 		\cdots		&	 c_1^{45}			\\
	\vdots		 &		\vdots			&		\vdots		&		\vdots			\\
	c_{N}^1 		 &		c_{N}^2 		   & 		\cdots		&	 c_{N}^{45}			\\
	\hline 		\\
	x^1c_1^1 	 &		x^2c_1^2 		& 		\cdots		&	 x^{45}c_1^{45}	\\
	\vdots		 &		\vdots			&		\vdots		&	\vdots				\\
	x^1c_{N}^1	 &    x^2c_{N}^2    	& 		\cdots		&   x^{45}c_{N}^{45}	\\
	\hline 		\\
	y^1c_1^1 	 &		y^2c_1^2 		& 		\cdots		&	 y^{45}c_1^{45}	\\
	\vdots		 &		\vdots			&		\vdots		&	\vdots				\\
	y^1c_{N}^1	 &		y^2c_{N}^2 	  	& 		\cdots		&	 y^{45}c_{N}^{45}
\end{pmatrix}
\cdot
\begin{pmatrix}
	\alpha^1 		\\
	\vdots			\\
	\vdots			\\
	\vdots			\\
	\alpha^{45} 	\\
\end{pmatrix}
=
\begin{pmatrix}
	P + B				\\
	\vdots			\\
	P + B				\\
	\hline 			\\
	M_y				\\
	\vdots			\\
	M_y 				\\
	\hline 			\\
	-M_x 				\\
	\vdots			\\
	-M_x
\end{pmatrix}
\label{peso}
\end{equation}
The system is divided in three blocks of $N$ equations: the first $N$ equations impose the weight force balance in each position (on the z axis); the following $2N$ equations impose the balance of torques along both the $x$ and $y$ axes. In total, this linear system has $3N$ equations and $45$ unknowns ($\alpha^i$).
To simplify the notation, let us write the system (\ref{peso}) as $\textbf{C}\alpha=\textbf{p}$.
Because of the lack of information about the torques ($M_y$, $M_x$) and the bias magnet force ($B$), additional sets of independent measurements were introduced.
A good method to increase the number of independent equations in the previous system, is to vary the weight of the mirror by adding a known amount.
Once this operation is completed, we obtain a system similar to the previous, compactly written as $\textbf{D}\alpha=\textbf{q}$. The driven forces will be different and will be noted with $d$ instead of $c$; the torques  will be different too, and  will be noted with $N_y$ and $N_x$. The total force amount will not be  $P+B$ anymore, but $P+B+Q$, where $Q$ represents the added weight. 
The weight variation was achieved with a special PVC annular support of $Q= (-0.289\pm 5\times 10^{-3} \rm)N$, with three little wedges, each having a 120 degrees of symmetry to provide a self-centering mechanism. This tool, once the mirror is removed, can be introduced from the top (see Figs: \ref{Gotto}, \ref{Schema}). The weight of the tool was chosen around $10\%$ of the weight of the shell to avoid any relevant stress on the glass around the inner ring. With this value, the weight variation can be considered homogeneous over the whole shell surface, and we don't expect a less accurate solution of the linear system for a small weight variation of the tool.

With these additional sets of measurements performed using the tool, it is possible to find $B$, $M_y$, $M_x$, $N_y$, $N_x$, and naturally the calibration constants $\alpha^i$ lining up the measurements \emph{without} and \emph{with} weight variation, and modifying the previous linear systems in order to add the unknown amounts to the solution vector:

\begin{equation}
-
\begin{pmatrix}
	c_1^1 	&	c_1^2 	& 	\cdots	&	c_1^{45}				&1		 &0      &0  	 &	0  	& 0 		\\
	\vdots	&	\vdots	&	\vdots	&	\vdots				&\vdots&\vdots &\vdots&\vdots	&\vdots	\\
	c_N^1		&	c_N^2		& 	\cdots	&	c_N^{45}				&1		 &0      &0     &	0  	& 0 		\\
	\hline 				
	x^1 c_1^1&	x^2 c_1^2&	\cdots	&	x^{45} c_1^{45}	&0     &1		&0     &	0  	& 0 		\\
	\vdots	&	\vdots	&	\vdots	&	\vdots				&\vdots&\vdots &\vdots&\vdots	&\vdots	\\
	x^1 c_N^1&	x^2 c_N^2& 	\cdots	&	x^{45} c_N^{45}	&0     &1		&0     &	0  	& 0 		\\
	\hline 		
	y^1 c_1^1	&	y^2 c_1^2	& 	\cdots	&  y^{45} c_1^{45}	&0     &0		&-1	 &	0  	& 0 		\\
	\vdots	&	\vdots	&	\vdots	&	\vdots				&\vdots&\vdots &\vdots&\vdots	&\vdots	\\
	y^1 c_N^1	&	y^2 c_N^2  & 	\cdots	&	y^{45} c_N^{45}   &0     &0 		&-1	 &	0  	& 0 		\\
	\hline
	\hline 				
	d_1^1 	&	d_1^2 	& 	\cdots	&	d_1^{45}				&1		 &0      &0  	 &	0  	& 0 		\\
	\vdots	&	\vdots	&	\vdots	&	\vdots				&\vdots&\vdots &\vdots&\vdots	&\vdots 	\\
	d_N^1		&	d_N^2		& 	\cdots	&	d_N^{45}				&1		 &0      &0   	 &	0  	& 0 		\\
	\hline 				
	x^1 d_1^1&	x^2 d_1^2&	\cdots	&	x^{45} d_1^{45}	&0     &0		&0  	 &	1 		& 0 		\\
	\vdots	&	\vdots	&	\vdots	&	\vdots				&\vdots&\vdots &\vdots&\vdots	&\vdots 	\\
	x^1 d_N^1&	x^2 d_N^2 & \cdots	&	x^{45} d_N^{45}	&0     &0		&0  	 &	1 		& 0 		\\
	\hline 		
	y^1 d_1^1&	y^2 d_1^2& 	\cdots	&  y^{45} d_1^{45}	&0     &0		&0   	 & 0  	&-1		\\
	\vdots	&	\vdots	&	\vdots	&	\vdots				&\vdots&\vdots &\vdots&\vdots	&\vdots 	\\
	y^1 d_N^1&	y^2 d_N^2& 	\cdots	&	y^{45} d_N^{45}   &0     &0 		&0   	 & 0  	&-1		\\
\end{pmatrix}
\cdot
\begin{pmatrix}
	\alpha^{1} 		\\
	\vdots			\\
	\vdots			\\
	\alpha^{45}	 	\\
	\hline
	B 					\\
	{M_y} 			\\
	{M_x} 			\\
	{N_y} 			\\
	{N_x} 			\\
\end{pmatrix}
=
\begin{pmatrix}
	P					\\
	\vdots			\\
	P 					\\
	\hline 			
	0					\\
	\vdots			\\
	0 					\\
	\hline 			
	0 					\\
	\vdots			\\
	0					\\
	\hline
	\hline      
	P+Q         	\\
	\vdots			\\
	P+Q 				\\
	\hline 			
	0					\\
	\vdots			\\
	0 					\\
	\hline 			
	0 					\\
	\vdots			\\
	0					\\
\end{pmatrix}
\label{delta-peso}
\end{equation}
To simplify the notation, let us write the linear system of eq. (\ref{delta-peso}) as $\textbf{E}\alpha'=\textbf{r}$.
So the solution vector will be composed of 50 elements: 45 calibration constants $\alpha^i$, one for each actuator; 1 total force of bias magnets $B$; 2 torques without added weight $M_y$ and $M_x$; 2 torques with added weight $N_y$ and $N_x$.
It is important to understand that if the rank of  the system $\textbf{E}\alpha'=\textbf{r}$ is maximum, we are able to find even the calibration constants, then the  torques and the bias magnet force.
\section {The proof code}
\label {proof}
The proof code is a simulation to numerically demonstrate that the previous systems are resolvable, written in IDL language (Research Systems Inc). 
The simulation is composed by several steps: in the first step, a simulated force pattern and a simulated $\alpha$ set is built. In the second step, using the previous simulated amounts, we resolve a system analogue to the $\textbf C\alpha=\textbf p$ system. In the last step we apply the same procedure on a system analogue to $\textbf E\alpha=\textbf r$. The details are treated below.

The demonstration was obtained building a simulated force pattern $\textbf F_s$ and a simulated set of $\alpha_s$. Dividing the lines of $F_s$ by $\alpha_s$, a matrix analogue to the $\textbf C$ matrix, named $\textbf C_s$, was built. The known term vector, $\textbf p_s$,  contains $P$ in the weight equations and $0$ in the torques equations. 
The goal is to demonstrate that, resolving the system $\textbf C_s\alpha=\textbf p_s$, we obtain an $\alpha$ set equal to $\alpha_s$ at least at the machine precision. 
Since the torques and bias magnets force are imposed to be 0, it is permitted to define at  most~$45-3=42$ independent columns for $\textbf F_s$. The last 3 columns are calculated directly by the system, that resolves for each line  equation~(\ref{equ}).

The system was successfully resolved using both the identity matrix of order 42 and a random $42\times42$ matrix:  $\alpha$  and $\alpha_s$ were the same at least at the machine precision. The 45 simulated $\alpha$ vectors were built using random numbers, supposing a 10\% normal scattering around 1. This  confirms that both $\textbf C\alpha=\textbf p$ and $\textbf D\alpha=\textbf q$ are resolvable.

The read out noise of the capacitive sensors relates to a distance error of $3\rm nm$ \emph{rms}. As the bigger strokes on the stiffness modes are typically less than $1\rm \mu m$ (and reach values around $1\rm \mu m$ only for some actuators in the tip-tilt modes), this noise was disregarded.

In analogy with the previous method, the demonstration that the simulated $\alpha$ are reproducible  were verified with a system, called $\textbf E_s\alpha=\textbf r_s$, which with the same procedure resolves numerically the  $\textbf E\alpha=\textbf r$ system. In this case, to generate the F patterns, the values of $B, M$ and $N$ (inserted in the solution vector) were initially assumed to be $0$.  
Given that these systems are overdimensioned, a Singular Value Decomposition method was successfully used to obtain a numerical solution\cite{Numerical_Recipes}.
\section{Laboratory results}
\label{lab}
Laboratory measurements permitted to derive the $\textbf c$ and $\textbf d$ vectors while recording the forces applied by the closed loop control system. This was performed after having driven, in sequence, 45 force  lineary independent patterns, equivalent to the eigenmodes of the stiffness matrix (stiffness modes of the mirror). 
The results are shown in figs. \ref{almir}, \ref{Aldi}: the values of the force calibrations are between $0.70$ and $1.1$ \emph{true~Newton~/~commanded~Newton} with errors between $6.6\times 10^{-4}$ and $1.6\times 10^{-2}$ \emph{true~Newton~/~commanded~Newton}.
The currents driven on the 45 actuators reproduce the
stiffness modes of the mirror. The forces measured on each actuator
for these 45 shapes, and the respective 45 negative
shapes, constitute a cycle of 90 measurements. Several cycles of 90
measurements were performed, in order to have  homogeneous data sets to calculate the error.
Then, the $B$, $M$ and $N$ values previously obtained were used with those groups of measurements in the $\textbf{C}\alpha=\textbf{p}$ system, permitting to  estimate the error by the RMS of the average on these $\alpha$ at the $3\sigma$ level.

The procedure to calculate the calibration constants is solid and fully automatic. However, other important aspects, like the variation of the calibration constants with the time and the temperature, deserve to be investigated. Different runs in different days and rough temperature measurements were performed, but the amount of these data is not enough to provide an efficient estimation for the error or the effects of temperature. During the several steps of the secondary adaptive mirrors development, these aspects will be faced. This will bring us to a completer characterization of the instrument.

\section{Conclusions}
\label{fine}
In the framework of the Large Binocular Telescope, the two secondary adaptive mirrors play a decisive step. The procedure described to increase the performances of the adaptive P45 prototype help to complete this step.
Experimental data on the force calibrations of the P45 suggest that it is possible to calibrate the system using an IDL code that acquires the current measurements driven to the actuators.
Calibrations are determined with an error smaller than 1.5\%. 
Furthermore, if an actuator, a magnet, or a coil of the actuator is substituted, it is possible to obtain a new calibration using the developed procedure.


\newpage

\section*{List of Figure Captions}

\begin{figure}[h!]
\centering
		{\includegraphics[width=8.4cm]{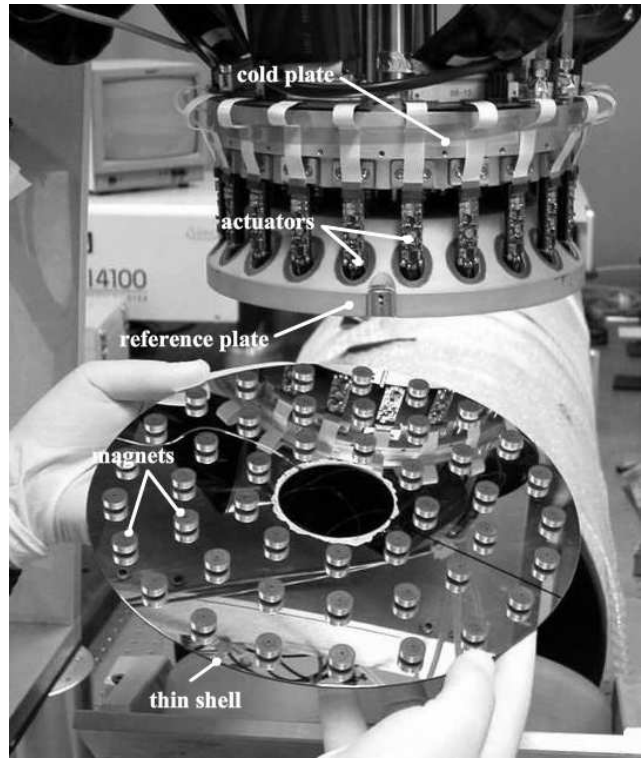}}
		\caption{Back surface of a thin deformable shell, magnets glued on its back, reference plate, actuators, and a cold plate. Riccardi-f1.eps}
		\label{p45}
\end{figure}

\noindent Fig. \ref{p45}. Back surface of a thin deformable shell, magnets glued on its back, reference plate, actuators, and a cold plate.

\begin{figure}[h!]
\centering
		{\includegraphics[width=8.4cm]{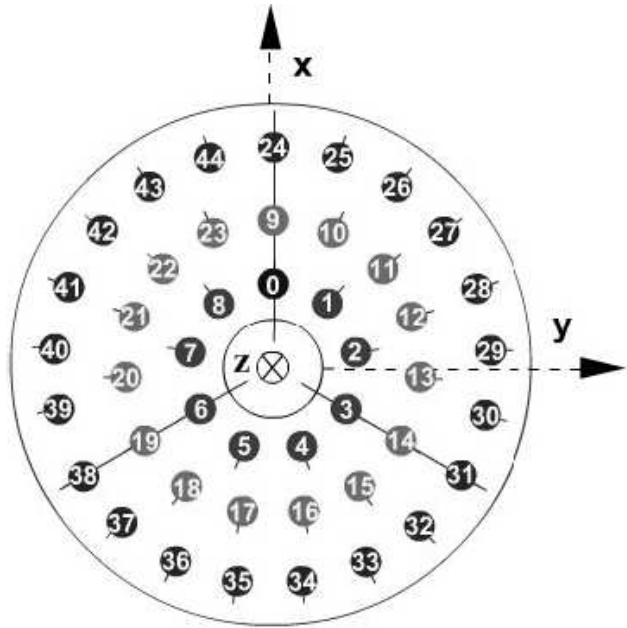}}
		\caption{Display of the actuators and $x$, $y$ axes. The gravity vector is along the $-z$ axis. Ricci-f2.eps}
		\label{Geometria}
\end{figure}

\noindent Fig. \ref{Geometria}. Display of the actuators and $x$, $y$ axes. The gravity vector is along the $-z$ axis. 

\begin{figure}[h!]
\centering
		{\includegraphics[width=8.4cm]{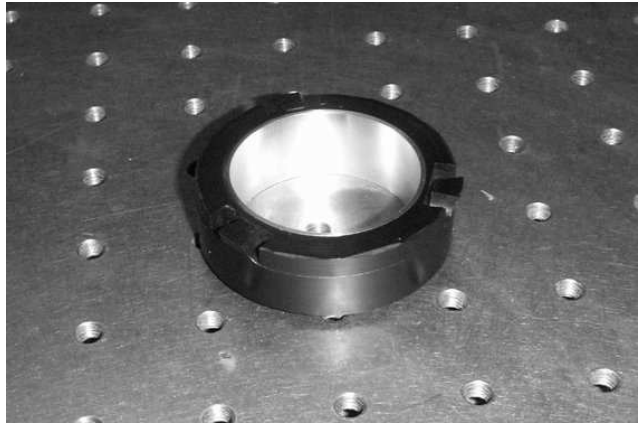}}
		\caption{Tool used to produce the weight variation. Ricci-f3.eps}
		\label{Gotto}
\end{figure}

\noindent Fig. \ref{Gotto}. Tool used to produce the weight variation.

\begin{figure}[h!]
\centering
		{\includegraphics[width=8.4cm]{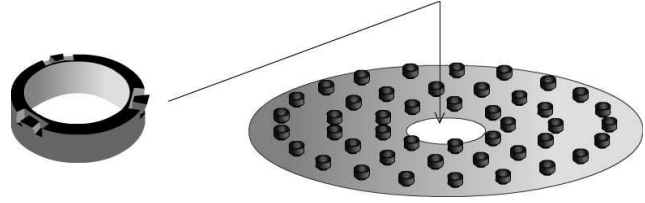}}
		\caption{Usage of the tool for the weight variation: the tool can be introduced from the back of the mirror previously removed from the structure. Ricci-f4.eps}
		\label{Schema}
\end{figure}

\noindent Fig. \ref{Schema}. Usage of the tool for the weight variation: the tool can be introduced from the back of the mirror previously removed from the structure.

\begin{figure}[h!]
\centering
		{\includegraphics[width=8.4cm]{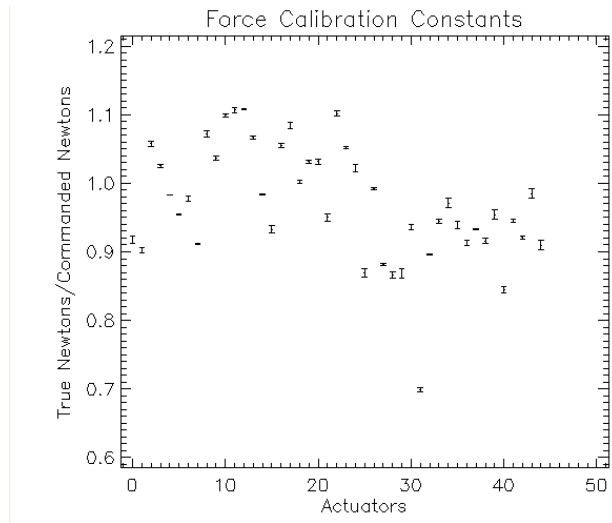}}
		\caption{Graphics of the force calibrations. Ricci-f5.eps}
		\label{almir}
\end{figure}

\noindent Fig. \ref{almir}. Graphics of the force calibrations

\begin{figure}[h!]
\centering
		{\includegraphics[width=8.4cm]{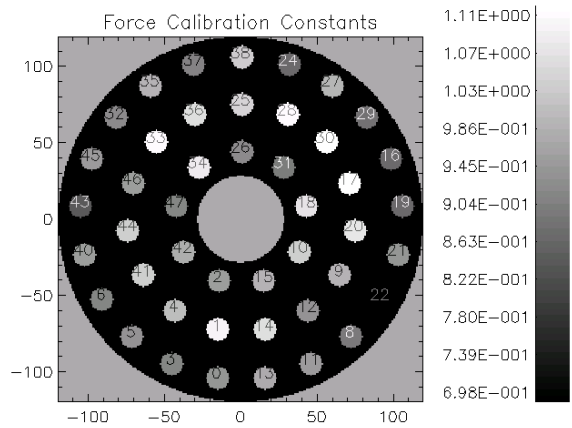}}
		\caption{Force calibrations on a display that illustrates the positions of the actuators. Ricci-f6.eps}
		\label{Aldi}
\end{figure}

\noindent Fig. \ref{Aldi}. Force calibrations on a display that illustrates the positions of the actuators

\newpage
\newpage
\newpage

\newpage

\newpage
\newpage

\end{document}